\documentstyle[epsf,prl,floats,aps]{revtex}

\newcommand{\be}{\begin{equation}}
         \newcommand{\ee}{\end{equation}}
         
\begin{document}
         \renewcommand{\topfraction}{0.99}
         \renewcommand{\bottomfraction}{0.99}
         \twocolumn[\hsize\textwidth\columnwidth\hsize\csname
         @twocolumnfalse\endcsname

         \title{Brane Gas Cosmology, M-theory and Little String Theory}
         \author{Stephon H.S Alexander}
         \footnote{E-mail: stephon@itp.stanford.edu}
         \address{~\\
         SLAC and Institute for Theoretical Physics,
         Stanford University, Stanford,CA 94305}
	 \maketitle

 \begin{abstract}
      We generalize the Brane Gas Cosmological Scenario to M-theory 
      degrees of freedom, namely $M5$ and $M2$ branes.  Without brane intersections, the Brandenberger Vafa(BV) arguments applied to M-theory degrees of freedom generically predict a large 6 dimensional spacetime.  We show that intersections of $M5$ and $M2$ branes can instead lead to a large 4 dimensional spacetime.  One dimensional intersections in 11D is related to (2,0) little strings (LST) on NS5 branes in type IIA.  The gas regime of membranes in M-theory corresponds to the thermodynamics of LST obtained from holography.  We propose a mechanism whereby LST living on the worldvolume of NS5 (M5)-branes wrapping a five dimensional torus, annihilate most efficiently in 3+1 dimensions leading to a large 3+1 dimensional spacetime.  We also show that this picture is consistent with the gas approximation in M-theory.                       
  \end{abstract}
 ]

\section{Introduction}
 
 The Brane Gas Cosmological scenario (BGC) provides a unification of 
 Cosmology with String Theory\cite{TseytlinVafa,Tseytlin,ABE,BrandenbergerEassonKimberly,Easson,Kabat,WatsonBrandenberger,matty}.  This is acheived by addressing the 
 dimensionality problem present in String theory by a dynamical 
 decompactification mechanism which involves the unwinding of annihilating 
 D-branes.   BGC also resolves the initial curvature singularities 
 which plague the standard big bang scenario by implementing 
T-duality of 
 the space-time as the radius of the Universe tends to zero\footnote{The issue of implementing T-duality for timelike singularities is still under investigation and is not as yet resolved.  In our case we assume that for each constant time slice one can map winding states on small radius to momentum states on large radius. \cite{Greg}}.  
 
 A key ingredient in BGC is that strings dominate the dynamics of the 
universe leading to decompactification of 3 large spatial 
dimensions.  The Brandenberger-Vafa (BV) mechanism which is central to BGC, posits that strings see each other most efficiently in $2(p+1)$ (p=1 for strings) dimensions\cite{TseytlinVafa,Tseytlin,bv}.  Therefore, their winding modes will annihilate most efficiently in $3+1$ dimensions and this leads to a decompactification in those dimensions.   The strings moving in higher dimensions eventually cease to interact efficiently and will fall out of equilibrium.   Their winding modes will prevent those higher dimensions from further expanding. 

 The BV mechanism was originally analyzed in the context of perturbative string theory.  We know that type II string theories at strong 
coupling becomes M-theory in 11 dimensions.  In this, $M5$ and $M2$ branes
play a fundamental role.    It is, therefore, important to reconsider 
the ubiquity of the BGC mechanism in all corners of moduli space 
and, in particular, the 11 dimensional corner where strings are not present as 
fundamental degrees of freedom\footnote{Brane Gases in 11D Supergravity was recently investigated by \cite{Kabat}.}.  We are also motivated to see how 
holography can play a role in describing the thermodynamics of the 
early universe.  What we will discover is that these two motivations 
are intimately connected and leads to a picture which  
dovetails the original proposal of Brandenberger and Vafa on string gases in the early universe.  

  The original paper on BGC examined a cosmological scenario where  
 a gas of strings and branes were wrapped over all cycles of $T^{9}$.  While further research has addressed 
 more general background topologies such as orbifolds and 
 inhomogenous space-times \cite{Easson,WatsonBrandenberger,EastherGreeneJackson}, no  
 procedure has yet emerged that generalizes brane gases in which 
the dynamics of $M2$ and $M5$ branes play a central role.  A related problem is that typical scenarios based on string theory contain a plethora of extended objects.  By reducing the dynamics to that of only two fundamental degrees of 
freedom, one hopes to 
obtain a clearer picture of the BGC mechanism.  In an 
earlier version of BGC, to evade the complications of dealing with 
complex configurations and interactions of brane gases, the authors 
of \cite{ABE}(ABE) only considered  parallel interactions of D-branes with the 
same $RR$ charges.   
 
 It is the purpose of this paper to initiate an investigation of some 
 of the above mentioned outstanding issues in BGC.  We will show that, by going 
 directly to the fundamental states of M-theory, the BGC scenario simplifies and also generalizes to 
 intersections and a strong coupling regime.  In this situation we 
 can use 11D supergravity and the worldvolume actions of $M2$ and 
$M5$ 
 branes to analyze BGC. 

\section{Brane Gases in M-theory:General Setting}
Brane Gases have been formulated in the weak string coupling regime 
in Type IIA where $g_{s} <1 $\cite{ABE}.  In this regime BGC was studied using 
IIA 
supergravity.  The zero mode fluctuation of p-brane winding and momentum 
states were obtained from linearized fluctuations of the Dira-Born-Infeld (DBI) action.
  In a toroidal geometry ABE \cite{ABE} were able to show that a hierarchy of dimensions decompactify resulting in three large spatial dimensions with two smaller extra dimensions.  However, the exact scaling of the hierarchy was 
not determined.  Can we gain more information about the scaling in dimensions by going to strong 
coupling?  

The strong coupling limit of $IIA$ is $11$ dimensional supergravity($SUGRA$), whose 
fundamental degrees of freedom are $M2$ and $M5$ branes.  We wish to 
study gases of intersecting $M2$ and $M5$ branes and generalize the 
nature of the intersection.  While our philosophy is rooted 
in the standard big bang scenario where the early universe existed in a 
hot dense state with all degrees of freedom excited, it is not a 
prediction of M-theory as to how extra spatial dimensions could start 
off small. 

Our starting point closely dovetails the standard hot big bang where 
all fundamental degrees of freedom are excited and in thermal 
equilibrium with all dimensions small and compact.  While we do not 
know what the quantum version of M-theory is, 11-D SUGRA is expected 
to be the classical limit.  We shall analyze our cosmology with an 
interplay between 11-D SUGRA and the effective worldvolume actions 
of $M2$ and $M5$ branes.  Let us start with the bosonic sector of 11D 
SUGRA.
\be S=\frac{1}{\kappa^{9}}\int d^{11}x\sqrt{-g}(R + F_{[4]}¥^{2} + 
F_{[4]}¥\wedge 
F_{[4]}¥\wedge A_{[3]}¥) \label{sugra} \ee
where $dA_{[3]}=F_{[4]}$ is a four form field strength tensor.

The dual of $F_{[4]}$ is a seven form which couples to the worldvolume of the fivebrane.  This state in 11-D is identified with the $M5$ brane, which is the magnetic dual of the $M2$ brane.  
The potential $A$ is a threeform and therefore couples to a $M2$ 
brane.  Now consider a gas of $M5$ and $M2$ branes which are not interacting with each other.  Since the five branes are heavier they will fall out of equilibrium faster than the $M2$ branes and their winding modes will cease to annihilate.  Membranes will dominate the dynamics at later times and according to the  BV topological arguments, 6 larger space-time dimensions will get big since the membrane winding modes annihilate most efficiently in these dimensions.  At this point, it is still not clear to realize three large spatial dimensions directly from M-theory dynamics.  To solve this problem we need to take brane intersections in M-theory into account.  
Indeed these intersections in M-theory are neccessary for 
consistency, such as local gauge invariance, which we will show later.  

The general setting starts with all branes and their intersections 
wrapping cycles of all dimensions small with respect to 
the string length scale.  In this case all 
branes wrap $T^{10,1}$ \footnote{ $T^{10,1}$ is a ten dimensional torus times the real time line  $T^{10}\times R^{1}$.} .  The equilibrium condition \be w_{p} + 
\bar{w}_{p} \rightleftharpoons loops, radiation \ee
states that there is an equal number of branes and anti-branes in 
the system interacting.  However, we expect the system to fall out 
of equilibrium when the winding degrees of freedom cease to 
interact.  
Moreover, the breakdown 
of thermal equlibrium is suggested by exceeding the Hagedorn 
temperature which is lower for higher dimensional
branes, so they are expected to fall out of equilibrium faster than 
lower dimensional branes.  
Hence we expect the five-branes to fall out of equilibrium faster 
than the membranes. 

We are working in 11 dimensional supergravity and there is no dilaton which in ten dimensional string theory, plays a key role to dynamically prevent cycles from expanding in the presence of winding modes. Therefore, in the 11 dimensional regime, it is necessary to  to show that branes wrapped on 
cycles will keep the dimensions from expanding; in other words there 
exists a confining potential for the modulus of the 
cycle.  The most straightforward way to show this is to work in a 
toroidal coordinate system that is time dependent in an 11-d SUGRA
background.  For simplicity we shall 
assume homogeneity of the background and adiabacity of the evolution.  We will discover that these assumptions are to weak to explain how membrane winding states can prevent expansion of the cycles they wrap.  Nonetheless, one will be able to isolate exactly where the assumptions break down.  The issue of the confining nature of wrapping membranes in M-theory was resolved by \cite{Kabat}.
 
We consider the wrapping of a membrane at rest along a two cycle.  We 
want to find its equation of state which can be obtained from finding 
the energy-momemtum tensor of the membrane in its wrapped 
configuration.
The relevant bosonic part of the membrane action is described by the 
Nambu-Goto action

\be T_{2}\int d^{3}\sigma \sqrt{-detg_{ab}} \ee
where $T_{2}$ is the tension of the membrane.  To find its energy 
momentum tensor with respect to the 11 dimension space-time we will 
vary the membrane action with respect to the background metric 
$G_{\mu\nu}$ defined as 
\be G_{\mu\nu}=g_{ab}X^{a}_{,\mu}X^{b}_{,\nu} \ee
The action is then
\be S_{2}= \int d^{11}x(-T_{2}\int d^{3}x \delta^{(11)} 
(X-X(\sigma))\sqrt{-\det{G_{\mu\nu}X^{\mu}_{,a}X^{\nu}_{,b}}})  \ee
Varying the action with respect to the background metric one obtains 
\footnote {see Boehm and Brandenberger for a more general treatment for DBI actions\cite{brandboehm}.} 
\be T^{\mu\nu} =-T_{2}\int d^{3}\sigma 
\delta^{(11)}(x^{\alpha}-X^{\alpha}(\sigma))\sqrt{-g}g^{ab}X^{\mu}_{,a}X^{\nu}_{,b} \label{energy} .\ee
from this we want to derive the pressure of the system of membranes 
wrapped on Tori.  This is obtained by averaging over the trace
$T^{ii}$ which is 
\be T^{i}_{i}(t,x)=T_{2}\int d^{2}\sigma 
\delta^{(11)}(x^{11}-X^{11}(t,\sigma^{i}))[3\dot{X}^{i}\dot{X}_{i}-3].\ee
where $\dot{X}^{i}\dot{X}_{i}$ is the squared velocity of a point on 
the membrane parameterized by the worldvolume coordinates.  By 
defining the mean squared velocity of the gas of membranes by 
averaging over all $\sigma^{i}$; i.e. 
$v^{2}=\frac{1}{d}\sum_{i}(\dot{X}^{i}\dot{X}_{i})$.
 It follows that the 
pressure can be obtained from the averaged trace.

\be \cal{P} \rm = \frac{1}{d}\sum_{i}T^{i}_{i} = [\frac{3}{11} v^{2}-\frac{p}{d}]\rho_{p} \ee
where $p$ is the WV spatial dimension and $d$ is the target space dimension, $\rho_{p}$ is the energy density of the membrane wrapped
around a p-cycle.
 
Since the membranes are at rest we are in the 
non-relativistic case $(v^{2} \rightarrow 0)$ and the pressure is, 
\be \cal{P} \rm =-\frac{1}{5}\rho_{p}\ee 
in agreement with the results of \cite{ABE,brandboehm}.  The pressure and the 
energy density will source the 11d Supergravity equations for the 
brane gas.

With this result in hand it remains to demonstrate that even in 11d 
without the dilaton, branes will keep the dimensions from expanding.
Similar to \cite{ABE} we will assume that the background space is a 
homogeneous and isotropic ten torus and that membranes and five branes can be 
wrapped over all of these cycles \footnote{The authors \cite{Kabat} have studied the more general anisotropic toroidal cosmological background.}  But for simplicity we shall focus 
for now on membranes.  The same argument will hold for fivebranes with 
exception to the difference in the equation of state for their 
winding modes.  Since we are mainly interested in the effects of a 
gas of membrane winding modes and transverse fluctuations on the 
evolution of a spatially homogenous Universe we will neglect the 
antisymmetric tensor fields.  We will use the metric of a torus and assume isotropy for simplicity,

\be ds^{2}= -dt^{2} + \sum a(t)dx_{i}^{2}. \ee

After varying the SUGRA action with this background and setting 
$\lambda{t}=log(a(t)) $ the resulting equations of motion are 

\be  d\dot{\lambda^{2}} = \frac{1}{2}e^{-d\lambda} 
P  \ee

\be d \ddot{\lambda}=-\frac{1}{2}e^{-d\lambda}(E+P)
\label{eom} \ee
where
\be E=\sum_{p} E^{nw}_{p} + E^{nw} \ee
\be P=\sum_{p} w_{p}E^{w}_{p} + wE^{nw} \ee
where the superscripts $w$ and $nw$ stand for the winding and 
nonwinding modes respectively.   
From eq [\ref{energy}] the energy $E_{2}$ of a membrane is 
\be T^{0}_{0 w} =T_{2}\int d^{3} \sigma \delta^{10}(x^{d} 
-X^{d})(t,\sigma)) \ee
and the total winding energy is
\be E_{2w} = T^{0}_{0}= T_{2}\int \rho_{2} = T_{2}vol_{p} \ee
and
\be Vol_{p}=a(t)^{p} \ee
we immediately see that the r.h.s of [\ref{eom}] acts like a potential for the radius of the torus.
\be V_{eff} \sim - e^{-d\lambda}(E+P) \ee
if the r.h.s is always negative, the potential is confining. 
In the case of dilaton gravity, the potential which appears instead of (18) determining the acceleration of the scale factor would be confining for negative pressures.  This is not the case here (for Einstein gravity).   The pressure of  membrane winding modes is always negative and will therefore by itself not prevent expansion of the cycle it wraps.  Therefore, at first sight it does not appear that wrapping membranes around tori in 11D supergravity will act to prevent expansion.  These conclusions, however, are based on the assumption of isotropy.  It is important to include anisotropic wrapping of the membranes.  Indeed, it was recently shown that eq [\ref{eom}] is generalzed to having a restoring force term with the inclusion of a wrapping matrix ,$N_{ij}$, which indexes the degree to which the membranes wrap the tori anisotropically.  The authors of \cite{Kabat} showed that at late times the anisotropic terms dominate the dynamics and act to confine the wrapped dimensions.  A very interesting case of wrapping matrix they considered, demonstrated that when three dimensions are fully unwrapped, four dimensions partially wrapped and three dimensions are fully wrapped, the three unwrapped dimensions will expand most rapidly.  The four partially unwrapped dimensions will grow slower than the three 'macroscopic' dimensions while the three fully wrapped dimensions will grow slowest.  Therefore there is a hierarchy of scale between the wrapped and unwrapped dimensions.  

 We will argue in the next two sections that the inclusion of string like intersections and turning on G-flux leads to the same late time decompactification behavior of a particlular  choice of wrapping matrix which gives rise to a hierarchy in scale between 3 macroscopic dimensions and two smaller dimensions\cite{Kabat}.

\section{Membrane and Five Brane Intersections}
As we pointed out in an earlier section, the mechanism to generate a 
large four dimensional spacetime would fail in 11 dimension if localized string 
like excitations were not present.  Hence, M-theory must consistently 
incorparate string like degrees of freedom from the outset. In what will follow we 
will provide a kinematical consideration for how string like states 
arise in M-theory in 11 dimensions. 

It is well known that $M2$ and $M5$ 
branes are not segregate entities and that worldvolume gauge 
invariance requires their 
mutual intersections for consistency.  From these 
considerations, we can see how string like states robustly arise.

The $M5$ brane action is given by \footnote{The kinetic term of the M5 WV 3-form, $h_{3}$, is actually zero because of self duality of $h_{3}$.  But we will use the $M5$ WV action of \cite{Bandos} and shall impose the self duality condition at the equation of motion level.}

\begin{eqnarray} S= &&-T_{M5}\int d\sigma{6}[\sqrt{-det(G_{ij}}) - \frac{1}{4} 
h_{[3]}\wedge 
*h_{[3]} -C_{[6]} \cr && + \frac{1}{2}C_{[3]}\wedge  db_{[2]}]. \label{five} \end{eqnarray}
where $dC_{[6]}=C_{[7]}$ is the magnetic dual of the G-flux $*G_{[4]}$, $b_{[2]}$ is the $M5$ WV 2-form; $h_{[3]}=db_{[2]} -C_{[3]}$ is the self-dual 3-form living on the M5 WV.  Notice that $C_{[3]}$ is the pull back of $A_{3}$ from eq \ref{sugra}, which supports the $M2$ brane. 

Notice the 2-form, $db_{[2]}$ in the second term of [\ref{five}].  This term suggests that a string like configuration couples to the five brane action.  
How do we confirm this?  Let us turn to the membrane world volume 
action. 

 In addition to the worldvolume term
\be \int_{M2}\sqrt{-detG_{ij}} \ee there is the coupling 
$\int_{M2}C_{[3]} $ to the 
three form field.  The membrane WV action is invariant under a local $U(1)$ gauge transformation 
\be C_{[3]} \rightarrow C_{[3]} + d\Lambda_{[2]} \ee
However, this only holds if the membrane has no boundary.  If the membrane has a boundary it is no longer invariant under the above gauge tranformation.  One then needs an additional field on the boundary whose gauge 
transformation will cancel the boundary contribution from the above 
piece.  The following modification does the job \cite{witten}.
\be \int_{M2}C_{[3]} -\int_{\partial M2}b_{[2]} \ee
where $b_{[2]}$ is a two form field living on the two dimensional 
boundary of the membrane worldvolume with the gauge transformation 
\be b_{[2]} \rightarrow b_{[2]} + \Lambda_{[2]} \ee.  But notice that this 
is the same field which lives at a two dimensional boundary on the 
five brane world volume.  One niavely would associate $db_{[2]}=h_{[3]}$ 
where $h_{[3]}$ is the five brane self dual field strength.  However 
this is not correct and we further need to make the latter gauge 
invaraint.  This was worked out to be 
\be h_{[3]}= db_{[2]} - C_{[3|W_{6}]} \ee 
Therefore as implied from the last term of  M5 WV action, $b_{[2]}$ is the source for the  localized string intersection of the $M2-M5$ branes and it is important that this field is turned on in describing BGC.  We shall discuss this issue in the next section.  This concludes our general argument as to why membranes will end on 
five branes on a string like intersection.

Now let us focus on the role of the $M2$ and $M5$ branes in this 
compactification in the context of brane gases.  Consider a gas of M-branes which can wrap all 
cycles of this product manifold.  Now the crucial ingredient is that 
$M2$ 
branes will intersect $M5$ branes on a string and these strings will 
also wrap the cycles admitted by the interstection rules.   $M2$ branes can also 
embed themselves into the $M5$ as well 
as intersect at a point. 

Given these string intersectons in 11D we arrive at a cosmological scenario where $M5$ branes will quickly annihilate in 10+1 dimensions.  However, the intersecting $M5-M2$ system will not fully annihilate due to the string intersection which wraps one cycles of $T^{10}$.  There are $M2$ branes with no boundaries and these will allow 5 spatial dimensions, $T^{5}$, to get large.  The string intersections that wrap the $T^{5}$ will fully annihilate in 3  spatial dimensions of this $T^{5}$.  Once the strings annihilate there will no longer be any obstructions for the fivebranes to fully annihilate.  Therefore, we end up with a hierarchy of dimensions where 3 dimensions have decompactified in a five dimensional manifold; which is identical to the conclusions of \cite{ABE} in the context of type IIA string theory.
In the sections that follow we provide some evidence for our conclusions.  

\section{The Role of Fluxes in Cycle Stabilization}
Initially we did not include the form fields present in 11D 
SUGRA, but they are expected to play an important role in the 
cosmological dynamics.  Here we provide an argument which demonstrates that G-flux can stabilize some of the moduli of a four torus.  More importantly, we want to show how M-theory encodes the breaking of 11 dimensional Poincare' invariance by turning on G-flux.  This leads dynamically to separation of a six dimensional torus $T^{6}$ from a four dimensional torus $T^{4}$. Furthermore,  we will also show in the following calculation that this G-flux is consistent with the initial conditions of BGC in M-theory namely that we have an equal number of membranes and anti-membranes.

  G-flux is a four form which couples to the five brane 
worldvolume via. its magnetic dual, a seven form $*G_{[4]}=C_{[7]}$.  
Also, G-flux can couple to the membrane intersection on the $M5$ brane as shown in the previous section by turning on the self-dual $M5$ three-form $h_{[3]}=db_{[2]}-C_{[3]}$; therefore membranes, can source G-flux.  We will show that membranes transverse to an eight dimensional torus will source G-flux which separates a four dimensional torus from a six dimensional torus,  dynamically breaking the 11 dimensional Poincare' invariance. 

 In the absence of membrane sources, the G-flux for a supersymmetric eight dimensional manifold (ie. CY4-fold) compactification of M-theory shoud vanish. But since we started with of an equal amount of membrane and anti membrane sources in 11-d we can explicitly show that the G-flux can be supported by the $T^{4}$¥manifold even though it is compact.  The Lagrangian for a membrane and anti-membrane is

\be \cal{L}\rm = \int G_{[4]}^{2} + n\int_{M2}C_{[3]} -m\int_{\bar{M}2} C_{[3]} + \int C_{[3]}\wedge G_{[4]}\wedge G_{[4]} \ee
where $m$ and $n$ denote the number of membrane and antimembranes respectively. Varying w.r.t $C_{[3]}$ leads to the equation of motion for the membrane source
\be d*G=n\delta^{8}(x) -m\delta^{8}(x) + G_{[4]}\wedge G_{[4]} \ee
yields the condition
\be m-n=  \int G_{[4]}\wedge G_{[4]} \label{gflux}\ee
whose solution is
\be \int G_{[4]}\wedge G_{[4]}=0 \label{vanish}\ee
if $m=n$, which corresponds to our initial condition in BGC that there are equal amount of membrane and antimembranes.  The right hand side of [\ref{gflux}] is easily satisfied provided that we identify a non vanishing $G_{6789} \neq 0$ with the coordinates on $T^{4}_{6789}$.  If the other components of the four form $G_{ijkl}=0$ vanishes then the condition in eq [\ref{vanish}] is easily satified.  Therefore we can have a non trivial G-flux which  factorizes the 11D Tori into a 
product space
\be T^{10,1} \rightarrow T^{5,1}\times T^{4}\times S^{1}\ee
due to the surviving G-flux.  From another point of view, we can 
similarly argue that a gas of $M5$ will wrap a  $T^{5,1}$ 
manifold and fall out of equilibrium faster than the $M2$ branes, 
leaving behind a $T^{4}$ that is supported by the $G$-flux 
\footnote{ $\int_{T_{4}} G_{4} $ the $G$ flux is supported by the four 
torus.}of other 
$M5$ branes which fell out of equilibrium.  

\section{Little Strings and Dynamical Decompactification}
We have argued that the BV mechanism necessitates string-like intersections in order to directly lead to large 4-dimensions.  Otherwise, the mechanism will generically predict 6 large dimensions.  We provided kinematical reasons which demonstrated that open membranes would lead to string intersections between $M2$ and 
$M5$ branes.  In the previous section we used the BV arguments to show that a hierarchy in scales between 3 larger space dimensions within 5 dimensions occur when string like intersections are taken in to account in M-theory.  This happens when the five branes wrap a $T^{4}$ while the other one dimensional WV direction which includes the stringlike intersection, wraps cycles of a transverse $T^{5,1}$.  Therefore a gas of five branes can all wrap $T^{4}$ and the string intersections on different five branes can all wrap the cycles of a tranverse $T^{5,1}$.  In this sections we provide another configuration which can lead to decompactification if the five branes wrap that whole $T^{5,1}$.  Moreover, this leads us to a handle of the thermodynamics of these string intersections in terms of holography.

 We do not have control of the microscopics of M-theory in 11-dimensions, but we want to gain thermodynamic 
information of the string like intersections between the $M2-M5$ branes.  We now explore a way of doing this i perturbative in the context of reducing M-theory to type IIA.  We will show that the string like intersections we are considering in M-theory is related to the Little Strings which occupy the WV of NS5 branes in type IIA when M-theory is compactified on a circle.   To summarize, the general picture we are considering 
at this stage:  One has a gas of spacetime filling $M5$ branes which have fallen out of 
equilibrium.  In other words we 
imagine a system of $N$ five branes with a gas of $M2$ branes ending 
on the $M5$s.   The five branes have ``condensed'' in the sense that they 
are all parallel and wrapping $T^{5,1}$.

Since we expect that these string like excitations in M-theory will dominate the late time dynamics of the universe leading to a decompactification to four space-time dimensions, we want to provide evidence in the type IIA picture which supports this geometrical picture.  But first we have to review how Little Strings in type IIA arise from M-theory.

Little strings are the IIA description of the intersection of M5 and M2 branes, where the latter stretches between a stack of five-branes.  The string like intersection reduces to little strings propigating on the WV of NS5 branes.  These theories (LST) are consistent non-gravitational theories in five and six 
space-time dimensions. LST can be derived from k $M5$ branes with a 
transverse circle of radius R in the limit of $R \rightarrow 0$, 
$M_{p} 
\rightarrow \infty$  with $RM^{3}_{p} =M_{s}^{2}$ kept constant.  With this connection between string like intersections in M-theory and LST in type IIA, we have a new insight into how type IIA can encode the dynamics of the string like intersection in M-theory.  We are specifically interested in understanding the BV argument for decompactification of the cycles that the gas of little strings wrap.  	

In the M-theory picture we developed in the earlier sections, membranes end on the world volume of the five brane.  This WV wraps $T^{5,1}=\cal{A}\rm$, therefore the string intersection will also wrap all cycles of $\cal{A}\rm$.   In order for  decompactification to occur on 
$\cal{A} \rm$,  the winding states of little strings in manifold $\cal{A}\rm$ 
have to be lighter compared to fundamental IIA strings in the bulk manifold $\cal{B}\rm=T^{4}$, 
$M_{s\cal{B}} \rm > M_{s\cal{A}} \rm$.   If this is case then the 
winding tension of strings will prevent $\cal{B} \rm$ from expanding 
at a slower rate than $\cal{A} \rm$.  
This is reminiscent of the construction of Easson where BGC was 
analyzed in the case of $K3$ compactifications.  In that case 
membranes wrapped both the $K3$ and $T^{4}$.  Easson conjectured that 
the membrane states were heavier in the $K3$ compared to $T^{4}$ 
however this was not proven\cite{Easson}.  In our case we have more information to 
analyze this situation since we have some control of LST at finite 
temperature.  

In order for the little strings to dominate the dynamics of the 
decompactification their energy per cycle on $T^{5,1}$ has to be less 
than the energy of fundamental strings on the $T^{4}$.
The tension of a membrane stretched between two five branes in 11d is
\be T_{M2}=M_{11}^{3} \ee
where $M_{11}$ is the eleven dimensional plank energy scale.  The little string which forms at the intersection has a tension
\be T_{ls}=M_{11}^{3}\times d\ee
where $d$ is the distance between the two fivebranes that connect 
the membrane.  We see that as the little strings become heavier as the five branes are further apart.  We need to establish the conditon such that the energy of the little string wrapped on the NS5 WV is less 
than  a wrapped string on $T^{4}$; this yields a consistency condition with the gas 
approximation.  The condition is:

\be M_{11}^{3}\times d \ll M_{11}^{3}\times R \ee
where R is the scale of a cycle of $T^{4}$.  This inequality leads to the condition
\be d<<R \ee
which means that the stack of parallel $M5$ branes are coincident compared to 
the size of the $T^{4}$ radius.  This suggests that we can treat the 
$M5$ worldvolume theory as a gas of little strings at finite density.  We will establish this in the following section.

\section{Thermodynamics of LST and the BV Mechanism}
In a previous paper we analyzed the brane gas scenario in the context 
of IIA SUGRA which can be related to the states of M-theory by 
compactifying on a circle.  In the last section we provided a 
consistency condition for the validity of the gas approximation in 
M-theory and found that there is  a thermodynamical 
realization of the little strings which are localized on a five 
dimensional submanifold $\cal{A}\rm$.  But how is one to know that we can treat 
this system perturbatively? The answer to this question comes from 
holography, similar in spirit to the AdS/CFT correspondence.  

Holography relates LST to string theory in the near horizon geometry 
of the fivebranes \cite{Aharony,abks,shiraz,gs}.   In general the theory is difficult to analyze 
since the near horizon geometry includes a linear dilaton direction, 
the real line $R$ along which the dilaton varies linearly
 
\be \Phi=-\frac{Q}{2}\phi, \ee
where $Q$ is a model dependent constant.  
Therefore the string coupling $exp{\Phi}=exp(-Q\phi/2)$ vanishes as 
$ \phi \rightarrow \infty$  which is the region far from the 
fivebranes and diverges as one approaches the fivebranes.  Thus, 
gaining holographic information of LST involves solving the dual 
theory at strong coupling.  However, there exists a  case where LST 
becomes weakly coupled and tractable by a perturbative holographic 
analysis; the high energy density regime.  This is a key realization 
because it tells us that we can understand the dynamics of the system 
of intersecting membranes in M-theory perturbatively via. holography. 
We are now in a position to use BV arguments for little strings wrapping the $T^{5}$ on the NS5 brane worldvolume.  This is possible because we can retrieve the thermodynamic information of little strings wrapped on $T^{5,1}$ from the holographic dual.  The energy-momentum of this gas,  will backreact on the background geometry due to its linear equation of state.  In what follows we shall retrieve the equation of state of a 
gas of little strings, which represent the intersection of $M2$ and 
$M5$ branes in the 11 dimensional description.

The authors of \cite{Kutasov} found that LST has a Hagedorn density of states at 
very high energies which follows from the little string partition sum
\be Z(\beta)=\int_{0}^{\infty} dE\rho (E) e^{-\beta E} \ee

\be \rho(E)= e^{S(E)} \simeq e^{\beta_{H} E} \ee

Also, finite energy corrections to the density of states were found 
with interesting high energy behavior

\be \rho(E)=E^{\alpha} e^{\beta_{H}} E[1 +O(\frac{1}{E})] \ee
where $\alpha $ is negative.  The density of states yields the  
temperature energy relation

\be \beta = \frac{\partial\rho}{\partial E} = \beta_{H} + 
\frac{\alpha}{E} \ee

Since $\alpha$ is negative the temperature is above the Hagedorn 
temperature $T_{H} =\frac{1}{\beta_{H}}$ and the specific heat is 
negative.  If one increases the temperature of the system the energy 
decreases.  This behavior has lead the authors to suggest the LST 
undergoes a phase transition above the Hagedorn transition similar to the case of fundamental strings in 10 D.  

 Below $T_{H}$, little strings 
will remain in thermal equilibruim.  By the BV arguments, the little strings will cease to interact in $d>4$, their winding modes will fall out of equilibrium and those cycles will no longer expand.  However, the little strings continue to maintain equilibrium in 3+1 dimensions by the usual topological intersection arguments discussed earlier, as the winding modes continue to annihilate, those 3+1 dimensions on $\cal{A}\rm$ will be selected to grow larger than the other cycles. 

The equation of state of the LS living on $T^{5,1}$ is
\be p=w_{a}\rho \ee
where
\be w_{a}=w_{w} + w_{m} \ee

the winding mode w parameter is

\be w_{w} = -\frac{p}{d} = -\frac{1}{5} \ee

This pressure and energy density of the little strings on $\cal{A}\rm$ will backreact on the background six dimensional geometry.  By the same arguments presented by ABE: The backreaction of the winding modes will generate an effective energy momemtum tensor which will couple to type IIA supergravity.  As these modes were shown to be lighter than the bulk string modes they will fall out of equilibrium last and allow 3+1 dimensions of the $T^{5}$ to get large \cite{ABE}.We outline the process as follows

\be T^{10,1} \rightarrow T_{LST}^{5,1}\times T^{4}_{G}\times S^{1} \rightarrow T^{3,1}_{large}\times T^{2} \times T^{4} \times S^{1} \label{large} \ee

\footnote {Note that the decompactification rate of $T^{3}$ within the $T^{5}$ depends on the degree to which the fivebrane winding energy also annihilates.  This is plausible since the little strings are sourced by five brane fields, so that when the little strings at high density annihilate, the fivebrane tension decreases on the cycles of $T^{3}$.}
This is precisely the late time behavior obtained by \cite{Kabat} for a specific choice of wrapping matrix.
 
\section{Discussion}

We have provided an 11-dimensional picture of Brane Gas Cosmology which takes into account $M2$, $M5$ branes and their intersections.  By studying their non-perturbative dynamics at finite density, we retrieve two consistent pictures in type IIA string theory.  The first picture involves the string intersections annihilating on a $T^{3}$ transverse to four dimensions which fivebranes wrap. The second picture corresponds to the type IIA reduction of M-theory on a circle, where a gas of little strings localized on the worldvolume of k coincident $NS5$ branes.  The thermodynamics of the (2,0) LST is well described via. holography and one is able to obtain the equation of state of the little string winding modes wrapped on tori.  Due to the backreaction of the gas of little string on the six spacetime dimensions that they occupy, the BV arguments leads to decompactification of 3+1 dimensions.  There also seems to exist the possibility that 6 dimensions can get large first and then little strings further allow 3+1 to get even larger.  It remains for future investigation to show explicitly that the fivebrane winding modes of the $T^{3}$ will decay as the little strings annihilate fully.  Nonetleless, the scaling arguments presented in this paper suggests that this expectation is plausible.
 
Furthermore, our inclusion of G-flux and stringlike intersection complements the investigation of \cite{Kabat} where anistropic wrappings were included in M-theory to show a hierarchy in scale of decompactified dimensions.  In fact a specific choice of wrapping matrix which leads to three large dimensions coincides with our analysis.   Recently a detailed description of Hydrodynamics of M-thery and holography was initiated.  We believe that these studies will be important for a deeper understanding of BGC in M-theory \cite{Herzog}.  

Notice that the final space-time in eq [\ref{large}] is the same as the result found by ABE \cite{ABE} where there is a hierarchy of sizes of decompactified dimensions.  Our result here is an improvement because the dynamics from the M-theory perspective goes directly from 11 D to 6 D by non-perturbave dynamics, namely the inclusion of G-flux.  Nontheless, a brane problem still persists but can be solved by having a loitering phase or inflation.  Loitering is a phase wherein the Hubble radius becomes larger than the spatial extent of the Universe, then there is no causal obstrustion for the winding modes of the membranes to annihilate.
 The fact that our assumptions of a gaseous regime in M-theory coincided with a holographic description of strings on NS5 branes deserves more investigation and it will be interesting to revisit and check explicitly some of the assumptions in terms of holography such as the unwinding mechanism of the little strings.  Another important question is whether there is a holographic solution of the brane problem. For example, ABE showed that a loitering phase could solve the brane problem.  We expect this mechanism to work since we are using the same effective actions for the back reaction of the little strings, nonetheless it is intriguing to understand if this mechanism is encoded in (2,0) holography.  We leave this for future investigations.   

\section{Acknowledgments}
I would like to thank Damien Easson and Gary Shiu for discussions on this project during its beginning stages.  I am especially grateful to Robert Brandenberger, Keshav Dasgupta, Michael Peskin and Mohammad Sheikh-Jabbari for reading and making useful suggestions of a rough draft of this paper.  I am also grateful to Ori Ganor, Shamit Kachru, Amir Kashani-Poor and Kelly Stelle for discussions and guidance throughout the project.  This research was supported by the DOE under the contract DE-AC03-76SF00515.  This paper is dedicated to the memory of Sonia Stanciu.


\end{document}